\documentclass[11pt]{article}
\usepackage{amsmath,amssymb,amsfonts,epsf}
\usepackage[nosort]{cite}
\usepackage[margin=1in]{geometry}

\usepackage{pslatex}

\newcommand{\fft}[2]{{\frac{#1}{#2}}}
\newcommand{\ft}[2]{{\textstyle\frac{#1}{#2}}}

\let\bm=\bibitem

\newcommand{\be}{\begin{equation}}
\newcommand{\ee}{\end{equation}}
\def\ba{\begin{array}}
\def\ea{\end{array}}
\def\ft#1#2{{\textstyle{\frac{\scriptstyle #1}{\scriptstyle #2}}}}
\def\fft#1#2{\frac{#1}{#2}}

\def\sst#1{{\scriptscriptstyle #1}}

\def\td{\tilde}

\def\dalemb#1#2{{\vbox{\hrule height .#2pt
        \hbox{\vrule width.#2pt height#1pt \kern#1pt
                \vrule width.#2pt}
        \hrule height.#2pt}}}

\newcommand{\bea}{\begin{eqnarray}}
\newcommand{\eea}{\end{eqnarray}}

\def\0{{\sst{(0)}}}
\def\1{{\sst{(1)}}}
\def\2{{\sst{(2)}}}
\def\3{{\sst{(3)}}}
\def\4{{\sst{(4)}}}
\def\5{{\sst{(5)}}}
\def\6{{\sst{(6)}}}
\def\7{{\sst{(7)}}}
\def\8{{\sst{(8)}}}

\begin{document}

\begin{center}\ \\ \vspace{0pt}
{\Large {\bf Escape Trajectories of Solar Sails and General Relativity}}\\ 
\vspace{10pt}
Roman Ya. Kezerashvili and Justin F. V\'azquez-Poritz

\vspace{5pt}
{\it Physics Department, New York City College of Technology, The City University of New York\\ 300 Jay Street, Brooklyn NY 11201, USA}

\vspace{5pt}
{\it The Graduate School and University Center, The City University of New York\\ 365 Fifth Avenue, New York NY 10016, USA}\\

\vspace{5pt}
{\tt rkezerashvili@citytech.cuny.edu}\ \ \ {\tt jvazquez-poritz@citytech.cuny.edu}
\end{center}

\vspace{10pt}
\noindent {\bf ABSTRACT}

\noindent General relativity can have a significant impact on the long-range escape trajectories of solar sails deployed near the sun. Spacetime curvature in the vicinity of the sun can cause a solar sail traveling from $0.01$ AU to $2550$ AU to be deflected by as much as one million kilometers, and should therefore be taken into account at the beginning of the mission. There are a number of smaller general relativistic effects, such as frame dragging due to the slow rotation of the sun which can cause a deflection of more than one thousand kilometers.\\

\noindent {\bf Keywords:}\ \ solar sails, general relativity, escape trajectories\\ 

\pagestyle{empty}

\noindent {\bf INTRODUCTION}

The exploration of the solar system's frontiers - the region between 50-1000 astronomical units (AU) from the sun - is a most ambitious and exciting technological challenge. Deep-space missions using
chemical propulsion are somewhat limited because they require a long duration, a high launch speed and an enormous amount of fuel. Solar sails are an alternative method of propulsion that could result in a cruise speed that enables the exploration of extra solar space during the span of a human lifetime, and may eventually be applied to interstellar exploration \cite{rk1,rk2}. A recent study \cite{rk7} shows that after sail deployments at parabolic orbit with 0.1 AU perihelion, a 937 m radius beryllium hollow body solar sail with a sail mass of 150 kg and a payload mass of 150 kg reaches 200 AU in 2.5 years, the sun's inner gravitational focus at 550 AU in about 6.5 years and the inner Oort Comet Cloud at 2,550 AU in 30 years.

A solar sail should be deployed as close to the sun as possible so that the force due to the solar radiation pressure (SRP) is maximized. This results in a cruising speed of 800 km/s or greater. In order to minimize the perihelion distance, it is necessary to use low density sail materials that are highly reflective and heat tolerant \cite{rk8,rk9,rk10}. In addition to these factors, the effects of curved spacetime in the region near the sun should be considered. In fact, the perihelion shift of Mercury, located at about 0.3 AU, was the first experimental verification of general relativity. Perihelion distances as small as 0.01 AU- 0.1 AU may be feasible for solar sails in the near future. The effects of curved spacetime on solar sails in bound orbits has recently been considered \cite{poritz1,poritz2}. Even though a solar sail in an escape trajectory is close to the sun for only a short time, perturbations to its motion during this period when the outward acceleration due to the SRP is greatest can translate into dramatic effects on long-range trajectories. 

Responding to an increasing demand for navigational accuracy, we consider a number of general relativistic effects on the escape trajectories of solar sails. For missions as far as 2,550 AU, these effects can deflect a solar sail by as much as one million kilometers. We distinguish between the effects of spacetime curvature and special relativistic kinematic effects. We also find that frame dragging due to the slow rotation of the sun can deflect a solar sail by more than one thousand kilometers.\\

\noindent {\bf 1.\ \ \ DEFLECTION DUE TO CURVED SPACETIME}\\

\noindent {\bf 1.1\ \ \ The Orbital Equations}

We begin by deriving the general relativistic orbital equations for an object traveling near the sun in the absence of the SRP. The exterior spacetime of the sun in the static approximation is described by the Schwarzschild metric:
\be\label{form}
ds^2=-f c^2dt^2+f^{-1} dr^2+r^2 \left( d\theta^2+\sin^2\theta\ d\phi^2\right)\,,\qquad f=1-\fft{2GM}{c^2r}\,,
\ee
where $r$ and $t$ are the heliocentric distance and time as measured by a distant static observer, respectively. Note that an Earth-bound observer at $r=1$ AU can essentially play the role of a distant observer. 

The 4-momentum of the solar sail is $p^{\mu}=m dx^{\mu}/d\tau$, where $x^{\mu}=(t,r,\theta,\phi)$ and $\tau$ is the proper time measured in the frame of reference of the solar sail. Spherical symmetry allows us to orient the coordinate system so that the orbit is confined to the equatorial plane at $\theta=\pi/2$, and thus $p_{\theta}=0$. Since the metric is independent of time and the azimuthal direction $\phi$, the corresponding components $p_t$ and $p_{\phi}$ of the 4-momentum are conserved. We define the constants of motion $E\equiv -p_t/m$ and $L\equiv p_{\phi}/m$, where $m$ is the rest mass of the solar sail. Thus,
\be\label{p1}
p^t =\fft{m E}{c^2f},\quad p^r=m\ \fft{dr}{d\tau},\quad p^{\theta}=0,\quad p^{\phi}=\fft{m}{r^2}L\,.
\ee
$p^2=-m^2c^2$ yields
\be\label{reqn1}
\left( \fft{dr}{d\tau}\right)^2=\fft{E^2}{c^2}-\left(c^2+\fft{L^2}{r^2}\right) f\,.
\ee
Differentiation of (\ref{reqn1}) with respect to $\tau$ gives the radial component of the 4-acceleration
\be\label{firstar}
a^r=\fft{d^2r}{d\tau^2}+\fft{GM}{r^2}-\fft{L^2}{r^3}+\fft{3GM L^2}{c^2r^4}\,.
\ee
Note that this can also be found by taking the covariant derivative of the velocity 4-vector.

We will now include the effects of the SRP. We assume that the backreaction of the electromagnetic radiation on the background geometry is negligible so that it acts on the solar sail only via the SRP.
Even though the coordinate $r$ does not measure the proper distance, the surface area of a sphere is still given by $4\pi r^2$. This means that the acceleration due to the SRP is given by the same expression as in the Newtonian approximation, which is 
\be\label{ar}
a^r=\fft{\kappa}{r^2}\,,\qquad \kappa\equiv \frac{\eta L_S}{2\pi c\sigma}\,,
\ee
where we are restricting ourselves to the case in which the surface of the solar sail is directly facing the sun. In (\ref{ar}), $\sigma$ is the mass per area of the solar sail, which is a key design parameter that determines the solar sail performance \cite{rk1,rk11,rk12}. Note that we will use values for $\sigma$ which are larger than that of the solar sail on its own by a factor of ten or more, since we are taking into account the mass of the load that is being transported. The coefficient $\eta$ represents the efficiency of the solar sail used to account for the finite reflectivity of the sail and the sail billowing. Typically the conservative value for the solar sail efficiency is $\eta =0.85$. The solar luminosity is $L_S=3.842\times 10^{26}\ W$ and the speed of light is $c=2.998\times 10^8$ m/s. In the Newtonian approximation, the radially outwards force due to the SRP effectively reduces the mass of the sun to be $\td M\equiv M-\kappa/G$, where $M=1.99\times 10^{30}$ kg is the sun's actual mass. However, we wish to emphasize that this effective renormalization of the sun's mass does not carry over to the general relativistic framework, since both $M$ and $\td M$ appear in the orbital equations of the solar sail.

Equating the expressions for $a^r$ given in (\ref{firstar}) and (\ref{ar}) and taking the first integral gives
\be\label{sailreqn}
\left( \fft{dr}{d\tau}\right)^2=\fft{E^2}{c^2}-c^2+\fft{2G\td M}{r}-\fft{L^2}{r^2}f\,.
\ee
From (\ref{sailreqn}) and the $\phi$ equation in (\ref{p1}), we find the orbital equation to be
\be\label{curvedorbital}
\left( \fft{dr}{d\phi}\right)^2=\left(\fft{E^2}{c^2}-c^2+\fft{2G\td M}{r}-\fft{L^2}{r^2}f\right) \fft{r^4}{L^2}\,.
\ee
\vspace{1pt}

\noindent {\bf 1.2\ \ \ The Deflection of Solar Sails}

Before the solar sail is deployed at the distance of closest approach $r=r_0$, the gravitational attraction of the sun causes the speed of the spaceship to increase as it gets closer to the sun. The fact that the Helios deep space probes traveled at the record speed of about 70 km/s at 0.3 AU enables us to extrapolate (using conservation of energy within the Newtonian approximation) that the following sampling of speeds $v_0$ are feasible for the near future: $v_0=133$ km/s at $r_0=0.1$ AU, $v_0=188$ km/s at $r_0=0.05$ AU, and $v_0=420$ km/s at $0.01$ AU.

From the metric (\ref{form}), we find that the proper time interval is related to the coordinate time interval by
\be\label{taut}
d\tau=dt \sqrt{f-\fft{1}{c^2f} \left( \fft{dr}{dt}\right)^2-\fft{r^2}{c^2} \left( \fft{d\phi}{dt}\right)^2}\,.
\ee
Using this, we can express the angular momentum parameter $L$ as
\be\label{L}
L=\fft{v_0r_0}{\sqrt{f_0-v_0^2/c^2}}\,,
\ee
where $f_0\equiv f|_{r=r_0}$. Since we are restricting ourselves to the case in which the force due to the SRP is purely in the radial direction, $L$ is still a conserved quantity. However, since the SRP is transferring energy to the solar sail, $E$ is no longer a conserved quantity. In particular, $E$ is the energy parameter of the solar sail at $r=r_0$ as measured by a distant observer. Since $dr/d\tau=0$ at $r=r_0$, we find that
\be\label{E}
E=c\sqrt{c^2-\fft{2G\td M}{r_0}+\fft{L^2}{r_0^2} f_0}\,.
\ee
\begin{figure}[ht]
   \epsfxsize=2.8in \centerline{\epsffile{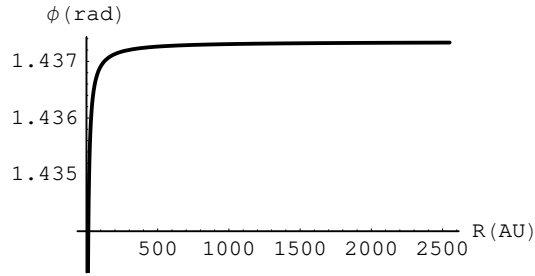}}
   \caption[FIG. \arabic{figure}.]{\footnotesize{\bf The angular position $\bf \phi$ versus the heliocentric distance $\bf R$ for a solar sail starting out at $\bf r_0=0.05$ AU with an initial speed of $\bf v_0=188$ km/s.}}
\label{fig6}
\end{figure}

From (\ref{curvedorbital}), we find that the angular position of the solar sail as a function of the heliocentric distance $R$ is given by
\be\label{h}
\phi = L \int_{r_0}^R \fft{dr}{r^2\sqrt{h}}\,,\qquad h\equiv 2G\td M \left( \fft{1}{r}-\fft{1}{r_0}\right)+L^2 \left( \fft{f_0}{r_0^2}-\fft{f}{r^2}\right)\,,
\ee
where we have taken $\phi=0$ at $r=r_0$. Note that $\phi$ can be expressed in terms of an elliptic integral of the first kind. Figure 1 shows $\phi$ versus $R$ for $r_0=0.05$ AU and $v_0=188$ km/s. Clearly most of the deflection of the solar sail occurs when it is in the vicinity of the sun.

In the Newtonian limit ($c\rightarrow\infty$), $\phi$ reduces to
\be
\phi_N=\cos^{-1} \left[ 1-\fft{v_0^2r_0^2}{G\td M}\left( \fft{1}{R}-\fft{1}{r_0}\right)\right]\,.
\ee
General relativity predicts that the solar sail will undergo a larger deflection than does the Newtonian approximation. Although the resulting difference in angle is rather small, this can translate into a large discrepancy $d=R(\phi-\phi_N)$ in the location of the solar sail for long-range missions. As shown by the left plot in Figure 2, $d$ dramatically increases for closer flybys, approaching as much as {\it one million} kilometers for a solar sail deployed at $r_0=0.01$ AU with $v_0=420$ km/s and traveling to $R=2550$ AU.

In order to disentangle the contribution to $d$ due to the kinematic effects of special relativity (within the Newtonian framework for gravity) from the effects of curved spacetime, we include the discrepancy between the special relativistic and Newtonian positions in the right plot of Figure 2 for the example of $v_0=420$ km/s at $r_0=0.01$ AU. While both types of effects are enhanced when the solar sail is deployed closer to the sun, it can be seen that the effects of curved spacetime dominate over those of special relativity.\\ 

\begin{figure}[ht]
\begin{center}
$\begin{array}{c@{\hspace{.1in}}c}
\epsfxsize=2.8in \epsffile{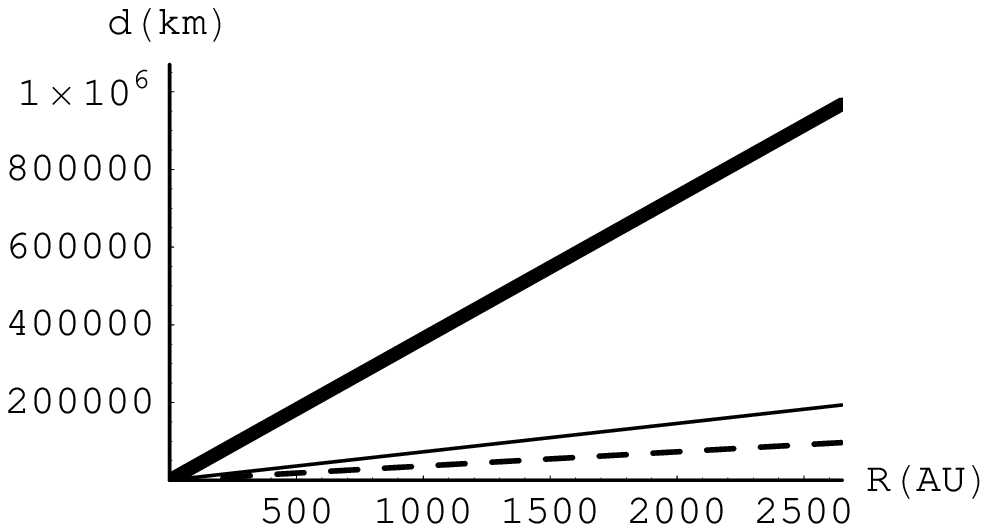} &
\epsfxsize=2.8in \epsffile{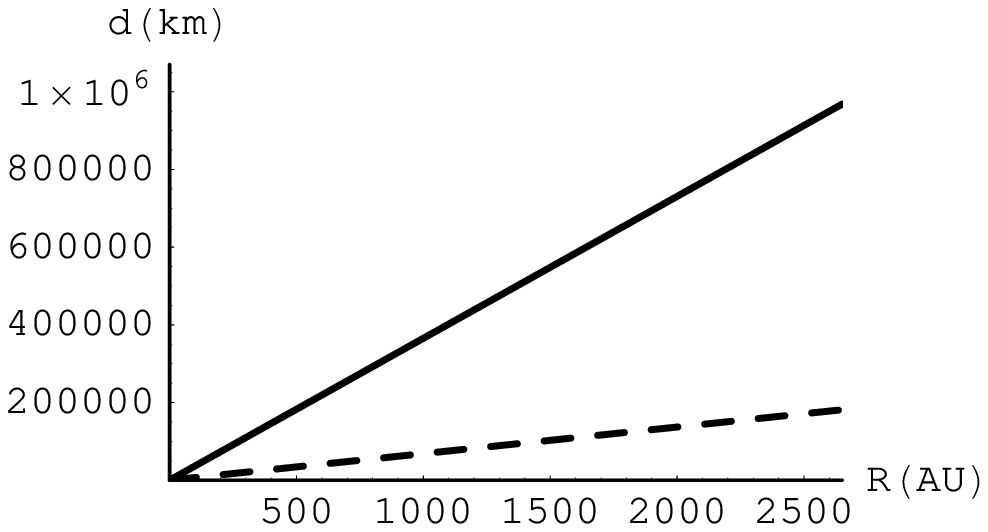}
\end{array}$
\end{center}
\caption[FIG. \arabic{figure}.]{\footnotesize{\bf The left plot shows the discrepancy $\bf d$ in the location of the solar sail versus the heliocentric distance $\bf R$ for the following sets of initial conditions: $\bf v_0=133$ km/s at $\bf r_0=0.1$ AU (dashed line), $\bf v_0=188$ km/s at $\bf r_0=0.05$ AU (regular line) and $\bf v_0=420$ km/s at $\bf r_0=0.01$ AU (bold line). The right plot shows the discrepancy in the location as predicted by special relativity versus Newtonian mechanics (dashed line) and general relativity versus Newtonian mechanics (regular line) for $\bf v_0=420$ km/s, $\bf r_0=0.01$ AU.}}
   \label{fig7}
\end{figure}

\noindent {\bf 2.\ \ \ OTHER EFFECTS OF CURVED SPACETIME}\\

\noindent {\bf 2.1\ \ \ Frame Dragging}

We will now consider the effect of frame dragging due to the slow rotation of the sun. The speed of the outer layer of the sun at its equator is $v\approx 2000$ m/s at the equatorial radius of $R\approx 7\times 10^8$ m. If we make the gross assumption that the core of the sun rotates with the same angular speed, then the angular momentum of the sun is given by $J=\ft25 Mv R\approx 10^{42}$ kg m$^2$/s. The external spacetime is described approximately by the large-distance limit of the Kerr metric \cite{kerr}
\be
ds^2=-f c^2 dt^2-\fft{4GJ}{c^2r}\sin^2\theta\ dt d\phi+\fft{dr^2}{f}+r^2 \left( d\theta^2+\sin^2\theta\ d\phi^2\right)\,.
\ee
We do not use the full Kerr metric since it does not seem to describe the external spacetime of a rotating material body, because it does not smoothly fit onto metrics which describe the interior region occupied by physical matter. 

We will restrict ourselves to trajectories that lie within the equatorial plane, for which the effect of frame dragging is maximized. Using perturbation techniques, we find the angular position of the solar sail can be expressed as
\be
\phi = L \int_{r_0}^R \fft{dr}{r^2\sqrt{h}}
\left[ 1+\fft{2GEJ}{c^4 L} \left( \fft{1}{fr}-\fft{1}{f_0r_0}+\fft{L^2}{hr^3}-\fft{L^2}{hr_0^3}-\fft{L^3}{Ev_0r_0^4}
\right)+\cdots \right]\,.
\ee
For a solar sail traveling from $r_0=0.01$ AU at $v_0=420$ km/s, frame dragging causes the location at $R=2550$ AU to be altered by approximately $1240$ km. The direction of the deflection depends on whether the solar sail is in a prograde or retrograde orbit relative to the rotation of the sun.\\

\noindent {\bf 2.2\ \ \ The Redshift Factor}

Besides deflection, the curvature of spacetime gives rise to a number of lesser effects, such as the slowing down of the passage of time near the sun. For example, an observer on Earth at $1$ AU measures about $31$ seconds more per year than does an observer at $r=0.01$ AU. This leads to a redshift in the wavelength of sunlight:
\be\label{redshift}
\fft{\lambda_{\infty}-\lambda}{\lambda}=\fft{1}{\sqrt{f}}-1\,,
\ee
where $\lambda$ is the wavelength measured by an observer at the heliocentric distance $r$ and 
$\lambda_{\infty}$ is the wavelength measured by a distant observer. It has been shown that the minimum thickness of the solar sail that provides maximum reflectivity depends on the wavelength of the solar radiation, as well as on the temperature \cite{rk11,rk12}. In particular, for fixed temperature, the optimum thickness of the solar sail increases with the wavelength. According to the redshift formula (\ref{redshift}), the wavelength $\lambda$ decreases as one gets closer to the sun where most of the acceleration occurs, which implies that the optimum thickness of the solar sail may also decrease. However, even at $r=0.01$ AU, the redshift is only $10^{-6}$, which has a negligible effect on the optimum thickness of the solar sail.\\

\noindent {\bf 2.3\ \ \ The Time Duration of a Voyage}

The proper time duration of a voyage in the reference frame of a solar sail traveling from $r=r_0$ to $R$ can be found from (\ref{sailreqn}), (\ref{L}) and (\ref{E}) to be
\be\label{deltat}
\Delta\tau=\int_{r_0}^R \fft{dr}{\sqrt{h}}\,,
\ee
where $h$ is given by (\ref{h}). This is generally less than the duration of the same voyage as measured by a distant observer (which can approximately be taken to be someone on Earth), which is
\be
\Delta t=\int_{r_0}^R \fft{dr}{\sqrt{f}}\ \sqrt{\fft{1}{c^2 f}+\left( 1+\fft{L^2}{c^2r^2}\right) \fft{1}{h}}\,.
\ee
For example, a 25 year-long voyage of a solar sail beginning from $r_0=0.01$ AU with $v_0=420$ km/s takes about $17$ minutes longer from the point of view of a distant observer.\\
\newpage

\noindent {\bf 2.4\ \ \ The Cruising Velocity}

The radial and tangential components of the velocity of the solar sail as measured by a distant observer at rest relative to the sun are given by
\be
v_r = \sqrt{f} \left[ \fft{1}{c^2 f}+\left( 1+\fft{L^2}{c^2r^2}\right) \fft{1}{h}\right]^{-1/2}\,,\qquad
v_{\phi} = L \fft{\sqrt{f}}{r} \left[ \fft{h}{c^2f}+1+\fft{L^2}{c^2r^2}\right]^{-1/2}\,.
\ee
For our example of a solar sail beginning at $r_0=0.01$ AU with $v_0=420$ km/s, the cruising velocity is about 480 km/s, almost entirely in the radial direction. While the tangential component is essentially the same as in the Newtonian approximation, the radial component of the velocity is faster by about 1.65 m/s, which is a difference that remains constant throughout most of the voyage and therefore has a cummulative effect.\\

\noindent {\bf CONCLUSIONS}

We have considered various general relativistic effects on long-range trajectories of solar sails. Small deviations in the initial trajectories of solar sails that are deployed near the sun can translate to large effects in the long run. For example, a solar sail deployed at $0.01$ AU can be deflected by one million kilometers by the time it gets to $2550$ AU. This deflection is primarily due to the curvature of spacetime near the sun, while the kinematic effects of special relativity contribute to a lesser degree. Frame dragging due to the slow rotation of the sun can result in a deflection of more than one thousand kilometers. A number of lesser effects of general relativity include the redshifting of sunlight, the slowing down of the passage of time near the sun, and a slightly increased radial component of the cruising velocity.


\end{document}